\documentclass[aps,amsmath,amssymb,superscriptaddress,twocolumn]{revtex4-2} 
\usepackage{graphicx}
\usepackage{comment}
\usepackage{amsmath} 
\usepackage{amssymb}
\usepackage{mathtools}
\usepackage{newtxtext} % Nuovo JFM
\usepackage{newtxmath} % Nuovo JFM
\usepackage{natbib}
\usepackage{hyperref}
\hypersetup{
    colorlinks = true,
    urlcolor   = blue,
    citecolor  = black,
}

\newcommand{\RomanNumeralCaps}[1]
\begin{document}

\title{Causally coherent structures in turbulent dynamical systems}

\author{Daniele Massaro}
\email{danmas@mit.edu}
\affiliation{Department of Mechanical Engineering, Massachusetts Institute of Technology, Cambridge 02139, Massachusetts, USA }
\author{Saleh Rezaeiravesh}
\affiliation{Department of Mechanical and Aerospace Engineering\char`,{} The University of Manchester\char`,{}  Manchester M139PL,{} United Kingdom}
\author{Philipp Schlatter}
\affiliation{Institute of Fluid Mechanics (LSTM)\char`,{} Friedrich--Alexander--Universität Erlangen--Nürnberg (FAU)\char`,{}  Erlangen 91058,{} Germany}

\begin{abstract}

The extraction of spatio-temporal coherence in high-dimensional, chaotic, non-linear dynamical systems, such as turbulent flows, remains a fundamental challenge in physics, mathematics and engineering. In this work, we employ Shannon transfer entropy (TE) to identify causally coherent motions in a zero-pressure-gradient turbulent boundary layer (TBL). This causality metric, rooted in information theory, enables the identification of sources and targets in dynamical systems using the corresponding time series. However, TE requires sophisticated tuning of various hyperparameters, such as the Markovian order of the source ($m$), which can spatially vary in wall-bounded turbulent flow. Here, we present an adaptive tuning and discuss the influence of $m$ across different TBLs. We introduce the concept of causally coherent structures (CCS), \emph{i.e.}\ coherent structures interpreted as spatio-temporal patterns of causality. Moreover, the net transfer entropy flux is also utilised to identify boundary layer locations acting either as sources or targets. The standard viscous, logarithmic, and outer layers are characterised by information fluxes, highlighting, for example, dominant top-down interactions between the inner and outer layers, analogously to the classical energy cascade. This work extends techniques previously employed in the literature, such as correlation and spectral analysis, and presents an approach that is inherently general and applicable to a wide range of chaotic dynamical systems, with applications in cognitive sciences, systems biology and finance.
 
\end{abstract}

\maketitle

Scientists have been identifying causal relationships in complex systems for centuries. In 1854, the physician John Snow mapped cholera cases in a complex system, as an urban environment, pinpointing a cluster in a contaminated water pump~\citep{snow1855}. His use of statistical metrics to address epidemiological uncertainty laid the groundwork for tackling causal relationships in complex systems with partially observable data. Nearly a century later, in 1948, Claude Shannon applied similar methods to communication uncertainty~\citep{sha1948}. Focusing on the transmission of signals in noisy environments, Shannon formulated the information-theoretic causality in complex dynamical systems. 
Turbulent flows are an example of such systems, from pollutant mixing in the atmospheric boundary layer to airflow dynamics in the human respiratory system. 

In contrast to idealized chaotic systems, \emph{e.g.}\ the Lorenz system \citep{Lorenz1963}, turbulent flows are high-dimensional, chaotic, non-linear dynamical systems. Thus, identifying causal relationships in such systems remains a significant challenge. In the past, both interventional \citep{jjmoin91,jjpinelli99,hwang2010,farrell2017} and non-interventional \citep{gonzalez,wollstadt21,harre,novelli} causality detection methods have been explored. The interventional approach is not straightforward, as it requires defining interventions that are both meaningful and non-disruptive, without modifying the system in a way that affects its interpretability. In addition, carrying out such interventions can be sometimes even impossible in experiments \citep{eberhardt2007,Eichler2013}. Alternatively, the non-interventional approach, also referred to as the non-intrusive approach, is based on a posteriori observations and is therefore well-suited to both experiments and simulations, without altering the dynamical system itself.

In this Letter, we extract causally coherent patterns based on the information-theoretic content of a turbulent dynamical system. We consider a zero-pressure-gradient turbulent boundary layer (ZPGTBL), a shear flow that develops near solid surfaces. Such boundary layer flows, common on aircraft wings and vehicle surfaces, are typically turbulent. Understanding mass and momentum transport in these layers is crucial for predicting aerodynamic loads and properties such as skin friction and heat transfer.
We pursue three key objectives.  \textit{i}) To explore the application of information-theoretic (IT) metrics for detecting causality in complex dynamical systems, highlighting their advantages over traditional correlation analysis. Correlation analysis has been a cornerstone in fluid dynamics and beyond, for example in astronomy, where it is fundamental to black hole imaging \citep{Klebanoff1973,Bagwell1993,Akiyama_2019}. However, it does not necessarily indicate causation. To overcome this limitation, more advanced measures such as Shannon transfer entropy (TE) have been developed. \textit{ii}) To assess the sensitivity of TE to critical hyperparameters, focusing on how these parameters vary across different system states, specifically as function of wall-normal distance and temporal history. \textit{iii}) To introduce a novel method to identify causally coherent structures (CCS) in the space-time parameter space of dynamical systems. Specifically, for TBL, CCS are educed across the boundary layers, quantifying their wall-normal asymmetry. Coherent structures (CS) are recurrent spatio-temporal patterns in turbulent flows, with great potential for flow control \citep{hussain86}. Similar patterns can be identified and extracted in other complex systems by considering an analogous parameter space, instead of space-time.

\begin{figure*}
\centering
\includegraphics[trim=0cm 1cm 1cm 0cm,clip,width=\textwidth]{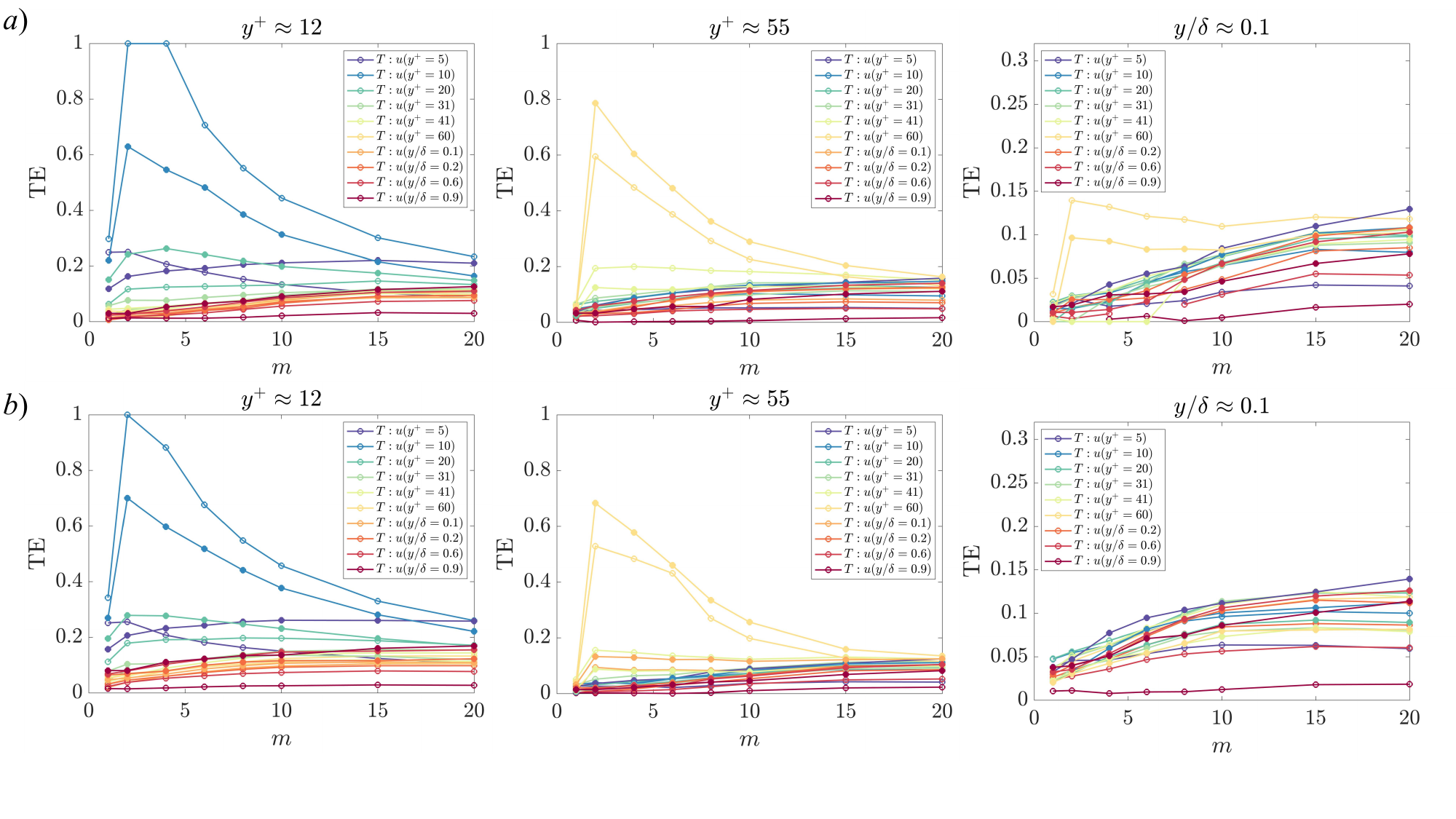}
\caption{The Shannon transfer entropy (TE) from, or to, the wall-normal reference locations as a function of the maximum source time lag, $m$. The panel \textit{a}) and \textit{b}) refer to $Re_\theta = 2240$ and $Re_\theta = 8000$, respectively. Circle-solid lines indicate cases where the streamwise velocity at the reference wall-normal location $y_{REF}=y_A$ is considered as the source, with the target specified in the legend; the reverse configuration is shown by filled-dotted lines.}
\label{fig:markord}
\end{figure*}

We begin by defining the adopted information-theoretical causality formulation. The Shannon entropy \citep{sha1948}, similarly to thermodynamic entropy by \citet{boltz}, serves as a macroscopic measure of disorder and uncertainty. By relating information flow to the one-way direction of time, \emph{i.e.}\ the time asymmetry \citep{Eddington}, it is possible to determine cause-effect relationships among processes in a system. Consider a continuous random variable~$Y$, which takes values~$y$ in a discrete (finite) set and has a probability density function (PDF)~$p(y) = \Pr(Y = y)$. Given an event $Y = y$, the information content of $Y$ is defined as: $I(y) = - \text{log}[p(y)]$, where the logarithm base is arbitrarily chosen. The Shannon entropy expresses the estimated mean value of $I(y)$~\citep{Shannon1949} and can be interpreted as a measure of the uncertainty in state $y$ \citep{wiener}. Considering two time series $X = x_t$ and $Y = y_t$, where~$x_t$ and~$y_t$ represent the source and target, \citet{sch2000} estimated the causality by computing the deviation from the condition of $x_t$-independence for the target $y_t$, using the generalized Markov condition:
\begin{equation}
p(y_{t}|\boldsymbol{y}_t^n,\boldsymbol{x}_t^m)=p(y_{t}|\boldsymbol{y}_t^n) \,,
\label{eq:markov}
\end{equation}
where $m$ and $n$ are the orders of the Markov process for the source~$x_t$ and target~$y_t$, respectively:
\begin{eqnarray}
\boldsymbol{y}_t^n&=&(y_{t-1},...,y_{t-n}),  \nonumber \\
\boldsymbol{x}_t^m&=&(x_{t-1},...,x_{t-m}),
\end{eqnarray}
with $m,n \geq 1$. The relation~(\ref{eq:markov}) is fully satisfied when the~$Y$-dynamics is independent of the present and past of~$X$. When this condition is not satisfied, the Kullback--Leibler divergence is used to measure the deviation
\begin{equation}
\text{TE}_{X \rightarrow Y}= \sum_{y_{t},\boldsymbol{y}_t^n,\boldsymbol{x}_t^m}  p(y_{t},\boldsymbol{y}_t^n,\boldsymbol{x}_t^m) \text{log}\frac{p(y_{t}|\boldsymbol{y}_t^n,\boldsymbol{x}_t^m)}{p(y_{t}|\boldsymbol{y}_t^n)} \,,
\label{eq:tecomplete}
\end{equation}
where $\text{TE}_{X \rightarrow Y}$ is the transfer entropy from $X$ to $Y$~\citep{sch2000}, with a lower bound of 0 (no causation). Among several existing normalizations, we adopt a conservative approach using raw TE values. 
Note that key aspects of TE are the direction and the robustness of the inferred information flow~\citep{report_madrid25}.

TE offers several advantages: it handles non-linearities (unlike Granger causality \citep{granger,wismuller2021}), implies directionality (distinguishing between  source and target), and accounts for history effects, considering the influence of the source over multiple past time lags. However, there are some limitations: challenging calibration  of hyperparameters, high computational cost required to reach statistical convergence, and the simplified bi-directional approach. To address these limitations, previous works have attempted to introduce a multivariate transfer entropy (mTE)~\citep{runge2012} or formulate novel definitions of causality, aiming to decompose the various co-causal components (synergistic, unique, and redundant decomposition of causality)~\citep{adriannature2024}. Exploring and comparing various causality definitions is out of the scope of this work. Instead, Shannon transfer entropy is applied to provide insights into causal relationships within turbulent flows. This approach is valuable not only for computational scientists, who typically have access to a complete dynamical system with billions of data points, but also for experimentalists, who may only have access to few time series measured by individual sensors in a physical system. In addition, the methodology is not limited to turbulent dynamical systems and is applicable to the time series of any complex system, ranging from brain neural networks~\citep{Wibral2014} to the stock market~\citep{stockmarket}.

Given any pair of time series, the IDTxl Python toolkit is used to compute TE~\citep{idtxl}. In this case, $x_{t}$ and $y_{t}$ are acquired from a ZPGTBL. Details of the numerical setup, grid resolution, and boundary and initial conditions can be found in \citet{eitel2014}. The Reynolds number ($Re$), \emph{i.e.}\ the ratio between the inertial and the viscous forces, varies between $Re_\theta \approx 2240$ to $8000$, based on momentum thickness ($\theta$), free-stream velocity ($U_\infty$) and kinematic viscosity ($\nu$). This is equivalent to friction Reynolds numbers of approximately $730$ and $2400$, based on the friction velocity ($u_\tau$) and boundary layer thickness ($\delta$). Here we focus on the lowest and highest $Re$. According to the classical layers interpretation of the turbulent velocity profile \citep{pope_2000}, we analyse time series of the streamwise velocity ($u$) at three distinct wall-normal locations: $y^+ \approx 12$ (viscous wall region, specifically the buffer layer), $y^+ \approx 55$ (logarithmic region) and $y/\delta \approx 0.1$ (outer layer, specifically wake region), where the plus units (with $^+$) denote the viscous scaling, and $\delta$ is the boundary layer thickness. Note that $y/\delta \approx 0.1$ corresponds $y^+ \approx 77$ and $y^+ \approx 248$, at $Re_\theta \approx 2240$ and $8000$, respectively. The three reference locations, hereafter named $y_{REF}=y_A$, are considered both as source and target. Time series were collected at specific spanwise locations ($z$). Given the statistical homogeneity in $z$, a physics-informed approach is followed, \emph{i.e.}\ spanwise averaging and $z$-symmetry are imposed. The Markovian orders introduced in equation~\ref{eq:markov} correspond to a physical lag of approximately $0.5$ time units ($0.64$ and $0.47$ at $Re_\theta \approx 2240$ to $8000$, respectively), expressed in inner units, \emph{i.e.}\ using the friction velocity and the viscous length scale.

The assessment of spatio-temporal coherence begins by evaluating how far into the past knowledge of the source can reduce uncertainty in the prediction of the target. Note that the target time lag is always considered $n=1$ as we aim to quantify the effect of the source history on the target present. Given the three wall-normal locations specified above, we observe that an universal source time lag ($m$) does not exist. At all three locations and both $Re$, the transfer entropy consistently peaks between $m=2$ and $m=4$ for the closest wall-normal position, see Figure \ref{fig:markord}. This result is crucial as it underscores the feasibility of defining a surrogate model for such time series based on a low-order models, specifically from second to fourth order. An examples are vector autoregressive models (VAR), a set of autoregressive models (ARMs) with preserving cross-covariance \citep{weibook}. VAR could also be used to predict near-wall dynamics by generating synthetic time series~\citep{ronith2024}. The findings align with previous measurements in turbulent channel flows \citep{massaro22caust,massaroentropy2023}, where the VAR order defined the system embedding time for the transfer entropy. In contrast, for the autocorrelation function, the system embedding time corresponds to the integral time scale of the time series~\citep{xavier2023}. This is also important for modelling and predicting time series of complex dynamical systems. However, as we move away from~$y_A$, whether~it acts as the source or the target, an increase in the source Markovian order ($m$) leads to an increment of TE, with a plateau around $m=15$. Beyond $m>20$, adding knowledge from the source no longer reduces the uncertainty of the target. A source Markovian order of $m=20$ appears to remain consistent across the various layers and Reynolds numbers. Particularly, in the viscous and logarithmic regions, convergence is observed at a lower order ($m\approx10$). 

Defining a reasonable $m$ is crucial, as the computational cost of TE increases linearly \citep{massaroentropy2023}. One possibility is to maximize the estimated TE \citep{ling2024}, without considering a time-history interval, but rather a single time lag. In contrast, our approach includes multiple time lags. Our findings do not yield a unique answer for $m$, suggesting an adaptive choice of $m$, which depends not only on the role of $y_A$ (either as source or target), but also on the wall-normal location. The choice of an adaptive $m$  as a function of the wall-normal location and source/target functionality is consistent with the physical interpretation of the Markovian order $m$ as a causal detector of streamwise velocity causal coherence, as discussed later. Further details are provided in the Supplemental Material.

\begin{figure*}
\centering
\includegraphics[trim=0cm 1cm 1cm 0cm,clip,width=\textwidth]{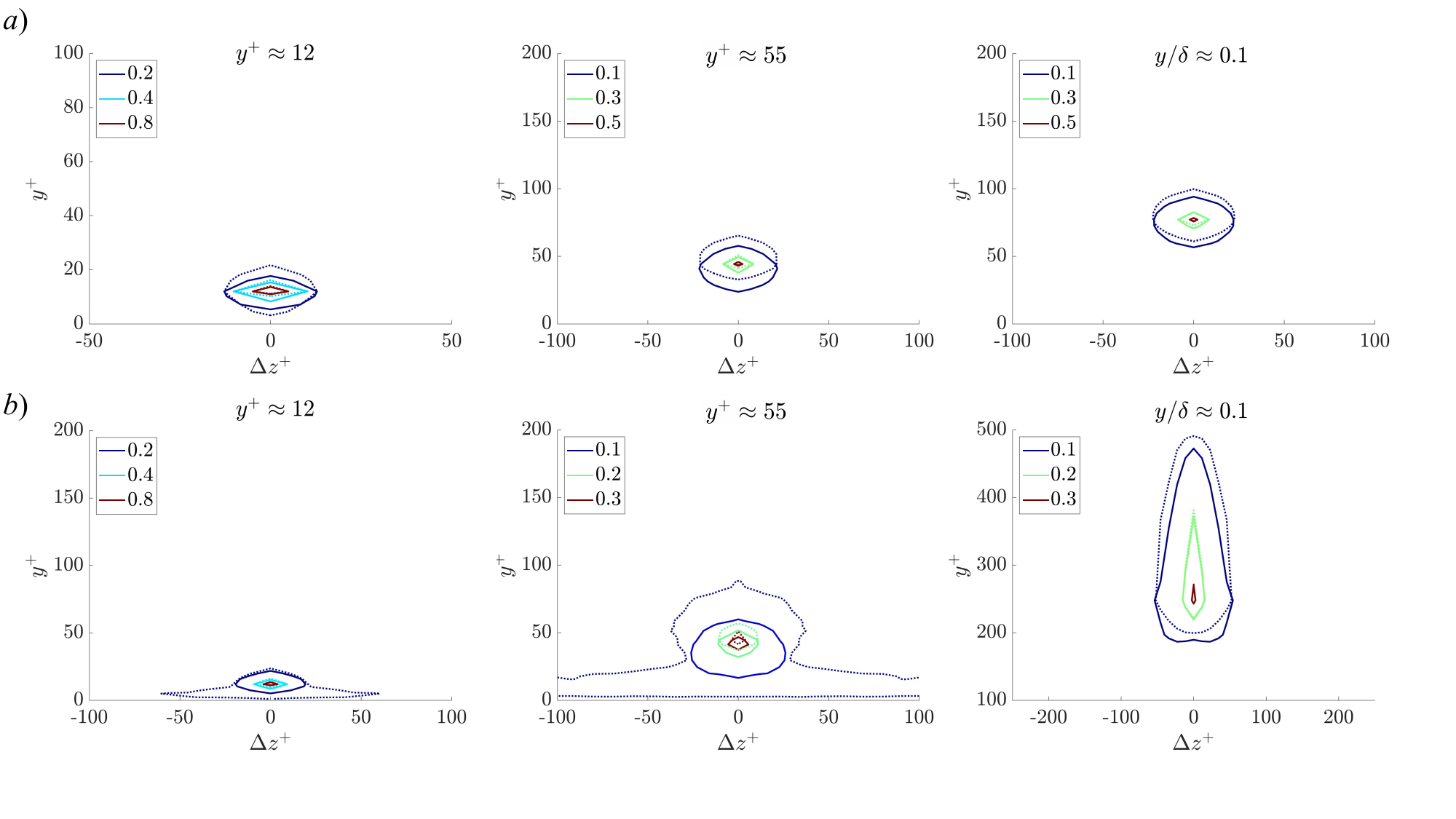}
\caption{Two-dimensional causally coherent structures of the streamwise velocity for a ZPGTBL at ($a$) $Re_\theta = 2240$ and ($b$) $Re_\theta = 8000$. From left to right, the wall-normal locations at $y^+ \approx 12$, $y^+ \approx 55$ and $y/\delta \approx 0.1$ are shown. In each panel, solid lines correspond to cases where the streamwise velocity at the reference point $y_A$ acts as the source, and dotted lines represent the reverse configuration. Three TE isolines are displayed; each enclosed region corresponds to TE values greater than or equal to the isoline value.}
\label{fig:TEmap}
\end{figure*}

The magnitude of TE merits a further discussion. In the viscous wall region, an almost complete causal relationship is observed for points in the neighbourhood of $y_A$, see left panels ($a,b$) in Figure \ref{fig:markord}. In the logarithmic region, at $y^+ \approx 55$ (where $y_A$ serves as the source), the TE magnitude is approximately halved (central panels) compared to $y^+ \approx 12$ (left panels). The reduction in TE magnitude arises from multiple factors, including interactions between the logarithmic layer and the inner and outer layers, as well as causation from alternative sources that this bi-directional approach may fail to detect. Remarkably, for $y^+ \approx 55$ (where $y_A$ serves as the target), the TE magnitude increases by 15\% w.r.t. the same location ($y^+ \approx 55$), where $y_A$ serves as the source. Thus, in the logarithmic region, the information from the area surrounding $y_A$ contributes more significantly to $y_A$ than vice versa, consistent with the view of an independent and $Re$ invariant logarithmic layer \citep{Townsend1976,antonia1982,smits2011,antonia2018}. In the outer region ($y/\delta = 0.1$), the increase in TE magnitude with $m$ reflects the slower temporal dynamics characteristic of large-scale structures. At the lowest Reynolds number, $y^+ \approx 60$ still serves as both source and target for small $m$. Indeed, $y/\delta = 0.1$ corresponds to $y^+ \approx 77$, implying limited spatial and scale separation. In contrast, at higher $Re$, a clear outer source of information is observed. For $y/\delta = 0.2, 0.6,$ and $0.9$, the information flows from the outer toward the inner region. Focusing on these three $y/\delta$ locations, the filled-dotted lines (where $y_A$ is the target and the source is specified in the legend) consistently exhibit larger values than the circle-solid lines; see the bottom-right panel of Figure~\ref{fig:markord}. As one may expect, large-scale motions at such locations, associated with the outer peak in the energy spectra at high $Re$~\citep{hutchinsmarusic2007,eitel2014,massaropodJFM}, have a strong causal influence on the inner layers. The analysis of such top-down causal interactions is presented in more detail later, when the net transfer entropy flux is introduced.

A careful balance between computational cost and the percentage of TE captured is essential. In this regard, it is encouraging that large values of $m$ are required to account for causality only from, or to, the viscous sublayer, whereas for $y^+ \geq 5$ smaller $m$ values are sufficient to capture most of the information. Thus, in line with the results discussed in Figure \ref{fig:markord}, we pick $m$ values between~4 and~15 for the remainder of the manuscript. Specifically, introducing the novel concept of CCS, we use $m = 8$. The idea of CCS extends correlation maps, which have previously been used to identify coherent patterns \citep{jimenez2018,taygun2022}. A three-dimensional correlation map for a boundary layer usually shows an inclined central positive lobe surrounded by smaller negative lobes. The cross-sectional areas exhibit near-circular shapes, whereas the streamwise sections appear elongated, see Figure 1 in \citet{sillero2014}. Note that the dimensions of the structures arising from correlation maps are much larger than those of individual structures \citep{jimenez2018,taygun2022}. Data from the turbulent boundary layer by \citet{eitel2014} have been used to generate the two-dimensional correlation maps in \citet{massaro2024phd}. Here, we reproduce the transfer entropy map by considering the three references $y$-locations ($y^+ \approx 12, y^+ \approx 55$ and $y/\delta \approx 0.1$), first as a source and then as a target, see Figure~\ref{fig:TEmap}.

\begin{figure*}
\centering
\includegraphics[trim=0cm 0cm 0cm 0cm,clip,width=1\textwidth]{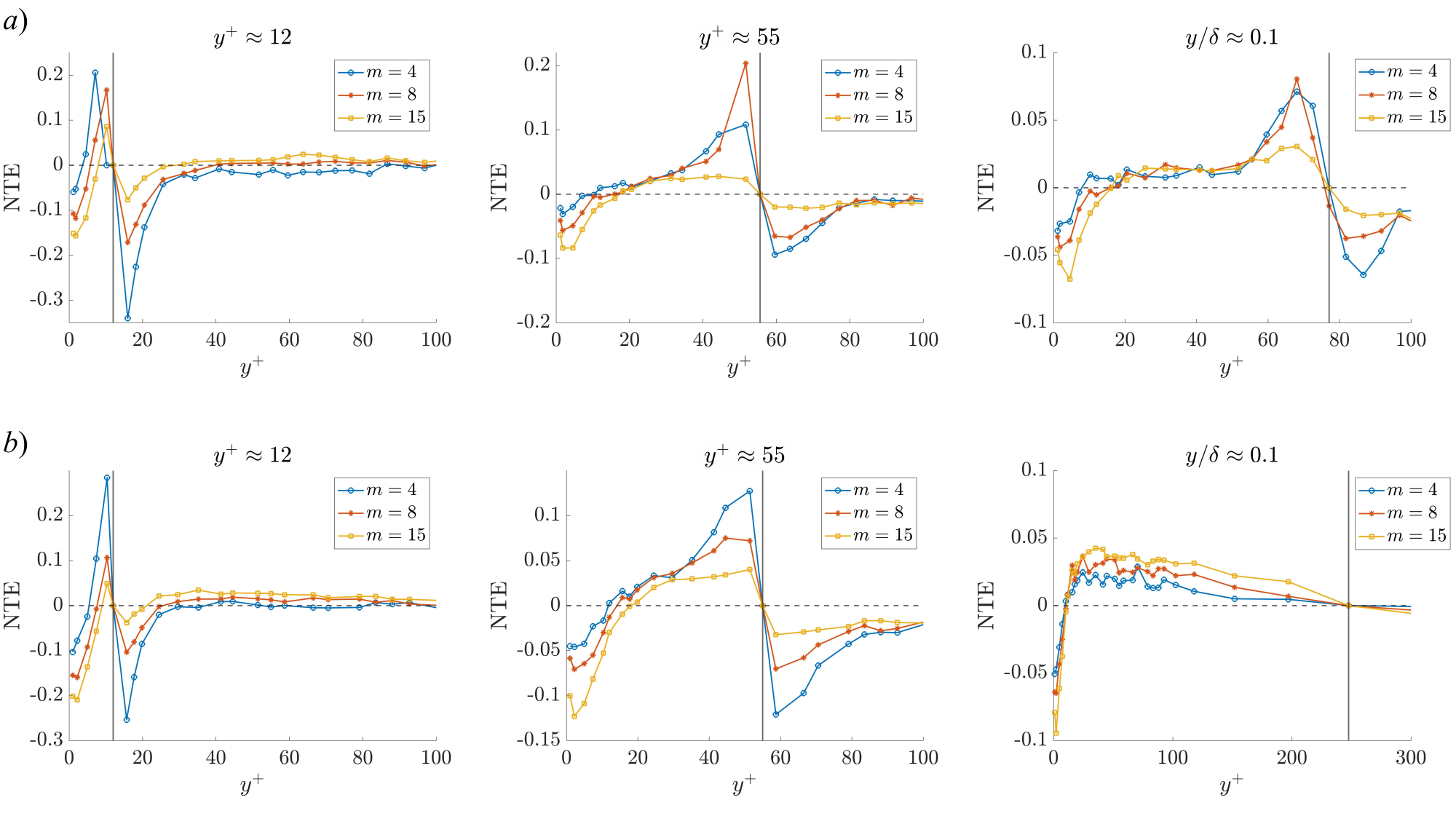}
\caption{The net Shannon transfer entropy flux (NTE) for a boundary layer at ($a$) $Re_\theta = 2240$ and ($b$) $Re_\theta = 8000$. From left to right, the wall-normal locations at $y^+ \approx 12$, $y^+ \approx 55$ and $y/\delta \approx 0.1$ are shown as reference wall-normal location, $y_A$.}
\label{fig:net}
\end{figure*}

At low Reynolds numbers, the educted CCS do not exhibit wall-normal asymmetry due to the limited separation between turbulent scales, see Figure~\ref{fig:TEmap}. The maps in panel ($a$) are symmetric with respect to the source and target, indicating no discernible direction for transfer entropy flux at various wall-normal locations ($y^+ = 12, y^+ \approx  55$ and $y/\delta = 0.1$). The TE magnitude shows that, in the near-wall region, most of the causality originates from the surrounding area, with TE reaching its peak just a few viscous lengths away from $y_A$. 
At the highest $Re$, where the separation of scales is more pronounced, a distinct behaviour emerges. Near the wall, at $y^+ = 12$, the dominant causal interactions still originate within a few viscous lengths of the reference location. The universality of small scales is corroborated here by the symmetry of information across various Reynolds numbers. However, a weak asymmetry appears already at $y^+ = 12$, see left panel $b$) in Figure \ref{fig:TEmap}. This asymmetry suggests an information flux from the wall to the logarithmic region. The trend is further confirmed at $y^+ = 55$, where the information flux originating from the wall and extending over a broader spanwise area, becomes more prominent. At $y^+ = 55$, the wall still imposes a causal influence, with no indication of a reverse effect. The TE magnitude decreases not only relative to that at  $y^+ = 12$, but also compared to the same location at lower $Re$, see the central column in Figure \ref{fig:TEmap}. Such CCS resemble cross-sectional views of the well-known velocity streaks~\citep{kline1967}, which play a central role in the autonomous near-wall cycle~\citep{jjpinelli99}. A three-dimensional analysis is required to clarify analogies with the typical inclination angles observed in correlation maps. According to the dynamic eddy interpretation by~\citet{jimenez2018}, CS are defined as strong flow features distinguished from the background via thresholding. Therefore, streaks are commonly extracted by applying thresholds to the fluctuating velocity component. Here, however, the full streamwise velocity field is considered. This typically yields irregular wall-attached layers, within which streaks appear as the very large uniform-momentum zones reported by~\citet{adrian2000} and~\citet{sillero2014}. Remarkably, the symmetric CCS shape changes in the outer layer ($y/\delta = 0.1$) where the wall effects, \emph{i.e.}\ the strong shear layer effects, are almost forgotten. The causality flows predominantly from the large-scale motions in the outer layer to the small-scale motions in the inner layers. The inner-layer motions are influenced by the outer layer, with top-down interactions prevailing. The causally coherent patterns extending upwards in the right panel $b$) of Figure~\ref{fig:TEmap} indicate such top-down interactions, in contrast to the left and central pictures. The interpretation of information-theoretic measures in fully turbulent flows, such as the ZPGTBL considered here, is considerably more complex than in simplified model flows, \emph{e.g.}\ the Kolmogorov flow used to study the transition to turbulence \citep{velaavila2024}. At least qualitatively, the structures identified here resemble the wall-detached eddies modeled by~\citet{perry1995} and \citet{marusic1995}. However, further investigation is required to confirm their statistical scaling and to explain why wall-attached structures, as described by~\citet{Townsend1951}, are not captured by Shannon transfer entropy. The presence of such top-down causal interactions agrees with previous studies reporting the modulation of near-wall turbulence by large-scale motions~\citep{hutchinsmarusic2007,mizuno2013,dogan2019,massaropodJFM,adriannature2024}. Further evidence is provided below introducing the net transfer entropy flux (NTE), see Figure~\ref{fig:net}.

NTE is drawn in analogy to the generalized Kolmogorov equation (GKE), which is an equation for the second-order structure function of the fluctuating velocity field, derived from Kolmogorov's theory \citep{kolmo1941,frisch1995,hill2002} and recently extended to wall-bounded flows \citep{cimarelli2013,zimmerman2022}. In this context, the hyperflux of scale energy in the compound space of scales enables the identification of forward/reverse cascades and descending/ascending spatial fluxes. For further details, the interested reader is referred to the relevant literature, see the appendix A in \citet{cimarelli2024}. Here, in analogy with the flux of scale energy \citep{cimarelli2024}, we define NTE at a given point $y_A$. NTE is estimated by considering $y_A$ as both a source (TE$_s$) and a target (TE$_t$) relative to the surrounding physical space. The net TE contribution is then evaluated as the difference (TE$_s$ - TE$_t$). For a portion $A$ of the space, where $\text{TE}_s - \text{TE}_t < 0$, NTE flows from the space $A$ to $y_A$, indicating that region $A$  exerts a stronger causal influence on $y_A$ than vice versa. According to the interpretation of TE by \citet{wiener}, this means the knowledge of the space $A$ reduces the uncertainty in understanding $y_A$. Considering the ZPGTBL data, the negative NTE extends up to approximately $y^+ \approx 12$ (averaged across various $m$ lags, see Equation~\ref{eq:markov}). Notably, this position aligns with the turbulent kinetic energy production peak~\citep{pope_2000}. At $y^+ \approx 12$, for all the considered wall-normal locations and for both Reynolds numbers, the net transfer entropy transitions from negative to positive. For the logarithmic and outer layers, at $Re_\theta=2240$ and $8000$, the interval $ 12 < y^+ < y_A^+$ exhibits positive net transfer entropy. This indicates that the streamwise velocity in this range is more causally-influenced by the velocity at the reference location $y_A$, rather than vice versa. On the contrary, for the region extending a few viscous units $y^+ > y_A^+$, the net flux becomes negative, implying that knowledge of this region would reduce uncertainty in the estimation of velocity at $y_A$. Consequently, across different wall-normal reference locations, the direction of the  \textit{causality arrow} shifts from top-down for $y^+ > 12$ to bottom-up for $y^+ < 12$. Since $y^+ \approx 12$ coincides with the peak in the production of turbulent kinetic energy, this transition is likely related to underlying energy-transfer mechanisms. However, further investigation is required. 
Examining the TE magnitude at both $Re$, NTE shows significant fluctuations at both $y_A^+ \approx 12$ and $y_A^+ \approx 55$. In these regions, the difference reaches up to $0.2$. At $y/\delta \approx 0.1$, the net transfer entropy is close to zero in the outer region. Particularly, at the highest Reynolds number, there is no increase/decrease across the reference location, indicating a NTE around zero; see right panel \textit{b} in Figure~\ref{fig:net}. Further analysis is needed to clarify whether this trend reflects a high Reynolds number effect or arises from insufficient velocity-probe resolution in the original dataset. However, in the inner space ($y < y_A$, with $y_A/\delta \approx 0.1$), the flux shows a causal influence from $y_A$ toward the surrounding area, whereas for $y > y_A$, the direction of causality is reversed, with a consistent top-down \textit{arrow of causality}. The positive NTE trend, beginning at $y^+ \approx 12$, extends up to the outer layer, where a positive NTE measures a flux towards the wall. These observations confirm the prevalence of top-down interactions, in alignment with the physical observation of top-down turbulent interactions  \citep{flack2005,mathis2009,chung2014,dogan2019}. This agreement is promising, as it indicates that NTE could support quantitative analysis of such interactions in more complex flows, where alternative approaches, such as the GKE, are computationally prohibitive.

The methodology used to extract CCS based on Shannon transfer entropy measurements is readily adaptable to different turbulent flows, or dynamical systems in general. Here, this causal metric is used to identify causally coherent patterns directly, rather than to evaluate interactions among pre-educed structures ~\citep{massaroNRP}, in a manner analogous to correlation maps. Two key differences distinguish CCS from classical correlation-based structures. First, CCS are typically smaller, offering a more realistic representation of the physical scale of the motions. Second, causality allows the identification of non-linear, asymmetric information flow, distinguishing sources from targets. A primary future direction involves exploring three-dimensional CCS in turbulent boundary layers, considering streamwise spatially developing structures. To address the computational cost of extracting such 3D motions, and associated with the PDF evaluation in the Shannon entropy formulation, alternative estimators may be introduced in the future, such as surrogate models capable of approximating Shannon entropy without explicitly calculating the PDFs \citep{Shao2014}. Moreover, the choice of the source time lag ($m$) is also critical. Indeed, a 3D causally coherent structure may exhibit space- and time-dependent variations with $m$.

In contrast to the present work, which analyses time series from the full dynamical system, an alternative approach involves constructing a reduced-order representation of the complex dynamical system and inferring bi-, or multi-variate, causality among the components of the network. Crucially, the effectiveness of this approach depends on the construction of the reduced-order model. In this sense, modal decomposition techniques, such as proper orthogonal decomposition \citep{lumley1967}, have proven to be a valuable path forward \citep{PODannrev,rowley2009,massaropodJFM}.

For turbulent flows, the concepts of CCS and net transfer entropy flux introduced here establish a connection between causality and the classic interpretation of the energy cascade. 
Extending the present methodology to additional flow quantities, such as other velocity components (either full or fluctuating), pressure fields, Reynolds stress tensor components, or vorticity, will provide a more comprehensive view of causal interactions. 
More broadly, regardless of the dynamical system under investigation, or its reduced-order representation, the discussed methodology enables the exploration of causal relationships using only two (enough long) time series. In the era of big data, fully data-driven metrics from different disciplines, such as information theory, are crucial to extract causally coherent patterns in the parameters space of a high-dimensional, non-linear system. This, in turn, facilitates understanding, controlling, and predicting various complex dynamical systems, from climate to neuro science.

 \begin{figure}
\centering
\includegraphics[trim=0cm 0cm 0cm 0cm,clip,width=0.5\textwidth]{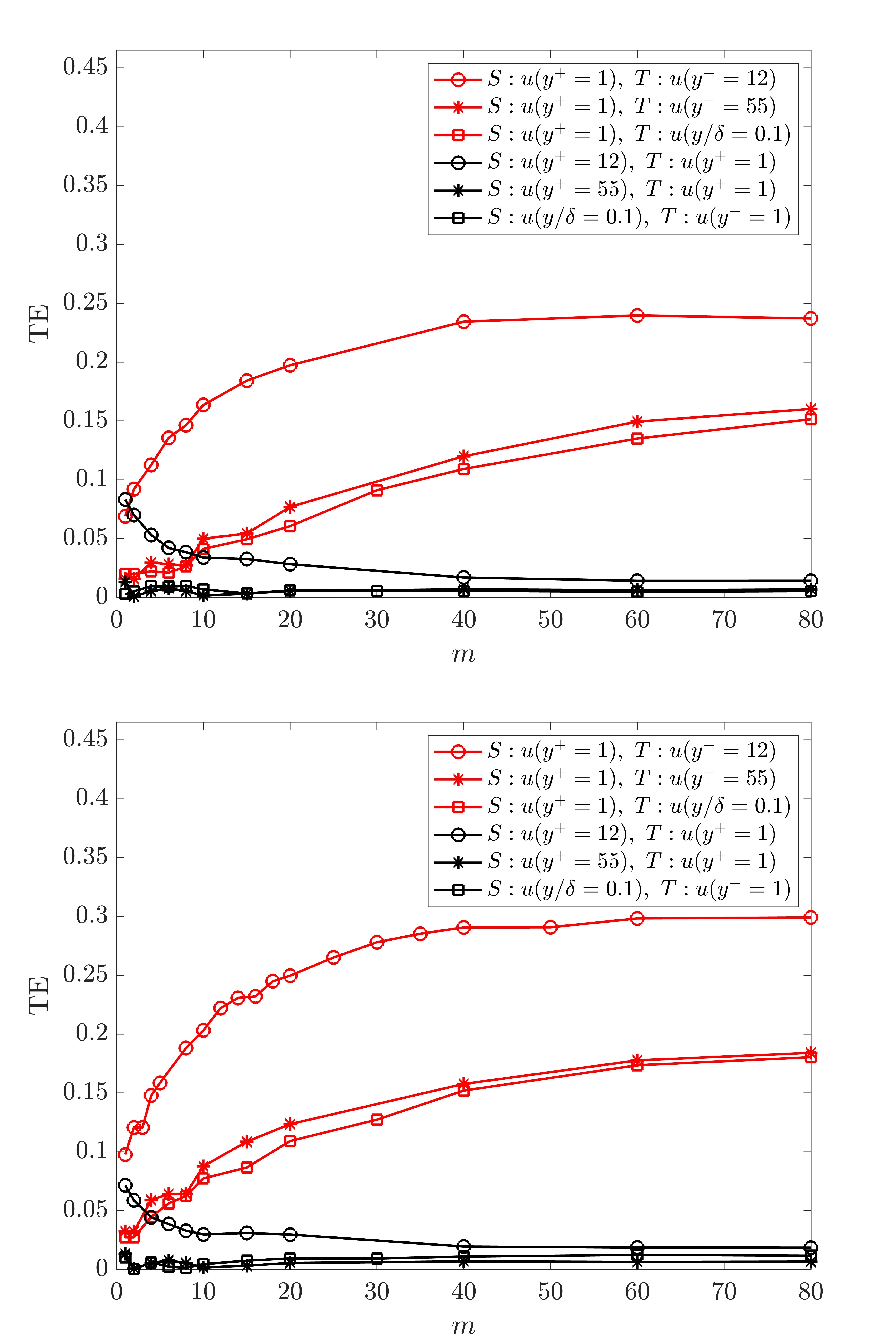}
\caption{The Shannon transfer entropy (TE) from, or to, the viscous sublayer (the streamwise velocity at $y^+ \approx 1$ is considered) w.r.t.\ the wall-normal reference locations. TE is expressed as a function of the maximum source time lag for (\textit{top}) $Re_\theta = 2240$ and (\textit{bottom}) $Re_\theta = 8000$.}
\label{fig:longtau}
\end{figure}

\section*{Supplemental Material}

Figure~\ref{fig:markord} in the main manuscript presents the wall-normal distances considered in this study. Here, the same analysis is extended to larger source time lags ($m$) at selected locations. For the three reference $y$-locations ($y^+ \approx 12, y^+ \approx 55$ and $y/\delta \approx 0.1$), the influence of $m$ was studied with respect to the viscous sublayer, which was observed to be the most critical region for TE convergence. This required extremely large lags, resulting in a significant computational effort. Figure~\ref{fig:longtau} shows that, from adjacent to the wall ($y^+ \approx 1$) to any $y_A$, the transfer entropy increases with the time lag $m$. As expected, the results do not differ substantially at the two Reynolds numbers, as the near-wall dynamics exhibit universal characteristics \citep{marusic2010,luchini2017,lee2024}. On the contrary, when $y_A$ serves as a source rather than a target, increasing $m$ results in a decrease in TE. This is attributed to the direction of information flux from the wall to the log layer. Figure~\ref{fig:longtau} also supports the chosen value of $m=8$ used in the TE maps, which represents a compromise between maximizing TE and maintaining reasonable computational cost.

The spanwise averaging of the resulting two-dimensional causal maps also deserves further comment. Indeed, the computation of TE is expensive, with cost increasing approximately linearly with $m$ \citep{massaroentropy2023}. Considering the three-dimensional nature of the flow, constructing two-dimensional, spatio-temporal causal maps is therefore extremely demanding. Consequently, the number of sampling points in the spanwise direction was limited to four. The corresponding TE values were then averaged, and symmetry was imposed, consistent with the statistically spanwise homogeneous nature of the ZPGTBL. To verify the influence of spanwise averaging, a prior critical assessment was performed using velocity probe measurements at various wall-normal locations with an increased number of sampling points in the spanwise direction.

\section*{ACKNOWLEDGMENT}

The simulations were performed on resources provided by the Swedish National Infrastructure for Computing (SNIC) at the PDC, KTH Stockholm. DM acknowledges the financial support provided by the Knut and Alice Wallenberg Foundation and the Karl Engvers Foundation for the generous travel grant that supported the research visit to the University of Manchester.

The authors report no conflict of interest.

\bibliographystyle{jfm}
\bibliography{thesis}
%Use of the above commands will create a bibliography using the .bib file. Shown below is a bibliography built from individual items.

\bibliographystyle{jfm}
%\bibliography{jfm2esam}

%% End of file `jfm2esam.bib'.

\end{document}